\documentstyle[pre,aps,floats,epsf,amssymb]{revtex}
\input epsf
\begin{document}
\draft

%\twocolumn[\hsize\textwidth\columnwidth\hsize\csname@twocolumnfalse%
%\endcsname
\title{Optimization of Green-Times at an Isolated Urban Crossroads  }
\vspace{1cm}

\author{M. Ebrahim Fouladvand$^{1,3}$ and Masoud Nematollahi$^{2}$  }

\address{$1$ Department of Physics, Zanjan University, P.O.
Box 313, Zanjan, Iran.}

\address{$2$ Tehran Traffic Control Company (TTCC), P.O. Box 15836  ,
Tehran, Iran. }

\address{$3$ Institute for Studies in Theoretical Physics and Mathematics
,P.O. Box 19395-5531, Tehran, Iran.}

\date{\today}
\maketitle

\def\ADD#1{{\bf{#1}}\marginpar{$\longleftarrow$ {\bf ADD!}}}

\newcommand{\bs}{\bigskip}
\newcommand{\be}{\begin{equation}}
\newcommand{\bel}[1]{\begin{equation}\label{#1}}
\newcommand{\ee}{\end{equation}}
\newcommand{\bea}{\begin{eqnarray}}
\newcommand{\ba}{\begin{array}}
\newcommand{\eea}{\end{eqnarray}}
\newcommand{\ea}{\end{array}}
\newcommand{\hfour}{\hspace*{4mm}}
\newcommand{\bra}[1]{\mbox{$\langle \, {#1}\, |$}}
\newcommand{\ket}[1]{\mbox{$| \, {#1}\, \rangle$}}
\newcommand{\exval}[1]{\mbox{$\langle \, {#1}\, \rangle$}}

%\begin{document}

%\twocolumn[\hsize\textwidth\columnwidth\hsize\csname@twocolumnfalse%
%\endcsname

%\title{Statistical Mechanics of an Urban Cross : A Solvable Model}
%                              %

%\date{\today}

%\maketitle

\begin{abstract}

We propose a model for the intersection of two urban streets. The traffic
status of the crossroads is controlled by a set of traffic lights which
periodically switch to red and green with a total period of$T$.
Two different types of crossroads are discussed. The first one describes the
intersection of two one-way streets, while the second type models the
intersection of a two-way street with an one-way street. We assume that the
vehicles approach the crossroads with constant rates in time which are taken
as the model parameters. We optimize the traffic flow
at the crossroads by minimizing  the total waiting time of the vehicles per
cycle of the traffic light. This leads to the determination of the
optimum green-time allocated to each phase.

\end{abstract}
]
%%%%%%%%%%%%%%%%%%%%%%%%%%%%%%%%%%%%%%%%%%%%%%%%%%%%%%%%%%%%%%%%%%%%%%%%

\section{Introduction}
Over the past decade, the vehicular traffic problems has been
intensively investigated within the context of statistical physics ( for a
review see Ref. \cite {css99,tgf97,tgf99,helbbook} ). Modeled as a system
of {\it interacting } particles driven far from equilibrium,
vehicular traffic presents the possibility to study various aspects of
truly non-equilibrium systems which are of current interest in statistical
physics \cite{zia,schutzbook,privbook}.
The majority of these studies have been allocated to the {\it highway traffic} 
. The other area in vehicular traffic is {\it urban traffic} which have 
also been studied by statistical physicists \cite{css99}. The simulation
of traffic flow in a large-sized city is a formidable task and many
degrees of freedom have to be involved (see e.g.
\cite{book,bell,robertson}). 
In practice, streets of a city form a network of junctions that are
linked together.  Each crossroads receives demands (vehicles
attempting to pass the cross ) and at each crossroads there exists a traffic
light which, with some certain programming, controls the transportation. 
The first model introduced by statistical physicists for the description of
 the city network, known as the BML model in the literature, uses
a deterministic cellular automata framework \cite{bml} and predicts a
sort of phase transition from free-flow to a jammed state. In the model,
each site of a square lattice represents the crossing of a
single-lane east-west street and a single-lane north-south street (no
turning of vehicles are allowed). The state of east-bound vehicles are
 updated synchronously at every odd
discrete time steps whereas those of the north-bound vehicles are updated
in parallel at every even time step following a rule which is simple
extension of totally asymmetric simple exclusion process (TASEP)
\cite{schutzbook,privbook}. \\

The BML model has been generalized to take
into account several realistic features of traffic in cities such as
asymmetric distribution of cars \cite{nagatani1}, faulty traffic lights
\cite{chung}, independent turning of the vehicles \cite{cuesta,nagatani2} 
and green-wave synchronization \cite{torok}. At first sight, the BML model
and the above-mentioned extensions seem unrealistic because the vehicles
hop from one crossing to the next. In a more realistic generalization,
each connecting bond of neighbouring junctions were replaced by a {\it
decorated bond } \cite{horiguchi,freund,chopard}. 
% but the model assumptions are too simple to be taken seriously for
%practical purposes.
Later, a more serious model of city network was introduced by
Chowdhury and Schadschneider (CS model) \cite{schad-deb} which developed
a more
detailed {\it fine-grained } description of city traffic. This model
combines the BML mode together with the Nagel-Schreckenberg model of
highway traffic \cite{NS,ito}. In the CS model, the signals are
synchronized in 
such a way that all the signals remain green for the east-bound
vehicles for a time interval $T$ and then, simultaneously turn red 
for the east-bound vehicles. To the best of our knowledge, The CS
model and its generalization \cite{brockfeld} is the most realistic model
introduced by the statistical
physicists for city network but nevertheless, despite its nice
formulation, there are still a
lot of simplifications which prevent it from being an {\it effective } and
applicable model to a city network.\\

In spite of developing models for
describing the city network, no detailed description of an {\it isolated
crossroads} has yet been explored. We strongly believe that in order to
have a better insight to the problem of city network, one must have a
clear picture at single crossroads which would definitely be of great
importance for an optimal programming of the entire network.
The empirical
mechanism by which the traffic lights are controlled
is generally divided into two distinct methods : fixed time and real-time.
In the fixed-time method, a fixed value of time is  allocated to the 
traffic light as well as its sub-phase times.  
 In the real-time method, which is becoming increasingly popular in great
cities, the
ensemble of crossroads are intelligently controlled by a central
controller.
The control mechanism  is usually based on the concept of
producing a kind of {\it green waves } between the crossroads.
 These waves interact with each other, and if the passing-demands are 
of high values, the green waves may have destructive effects on each
other, and hence, the concept of green wave may fail to be the final
solution for optimizing the overall flow. Knowing the local optimum
behaviour of single intersections could give us an appropriate criterion
to control the set of crossroads in an adaptive manner. Additionally,
not all the intersections of a city are highly affected by the
neighbouring intersections. To a good approximation, marginal
intersections could be regarded as isolated crossroads and therefore 
single-crossroads optimization strategies need to be investigated. 
In this paper
we aim to analyze a single crossroads in detail in order to find a better
insight to the problem of optimizing the total flow in cities.\\ 

The organization of this paper is
as follows: In section two, we introduce the model, state our
strategy for optimizing the traffic flow at the {\it one-way to one-way}
intersection and finally obtain the optimum green-times of the
corresponding phases. In section three, we extent our model to allow
for a three-phase {\it one-way to two-way } crossroads. Section four
is devoted to some empirical data on two of the Tehran crossroads which
are
under the control of an {\it intelligent traffic controller system}. We
compare the averaged green-time proposed by system to the optimum
green-time of our fixed time theory. Our theory input-parameters are
obtained via the empirical data. Finally we give our concluding remarks in
 section five.

\section{ Formulation of the Model}
\subsection{One-Way to One-Way Crossroads}
Let us consider a single crossroads which is the result of the
intersection of two perpendicular
 streets. In
their simplest structure, these streets can each direct a one-way traffic
flow. With no loss of generality, we take them as one-way South to North
(S-N)
and West to East (W-E) streets. Cars arrive at the south and the west 
entrances of the crossroads. In our model, we assume that the arrival
rates of
the cars, i.e., the number of cars reaching the crossroads per
second, are
constant in time. Although everyday driving experiences in cities
indicates that these rates have inevitable fluctuations in the course of
time, yet
in definite time intervals, the assumption of the constant arrival
rates could be justified at least on an average level. As will be seen
in what follows, this assumption leads to great simplifications. We take
the arrival
rates to be $\alpha _1$ (for S-N cars ) and $\alpha _2$ (for the W-E
cars) respectively. Also we denote the passing-rate of cars ( number of
cars passing the crossroads in the unit of time during the green-phase ) by
$\beta_1$ and $\beta_2$. The period of the traffic lights is taken to be
a definite value $T$ which is assumed to remain constant. The starting time
 of each cycle of the traffic light is the
moment at which the light turns green for the S-N street. The S-N light
remains green for $T_1$ seconds. At $T_1$ the traffic lights turn
red for the S-N street and simultaneously changes to green for the
W-E street. This is the beginning of the second phase which continues from
$T_1$ to $T$ (end of the cycle). During Phase I ( $ 0 \leq t \leq T_1$
), the S-N cars can pass the crossroads northwards and W-E cars are
stopped for
the red light.  Over Phase II ( $T_1 \leq t \leq T$ ), the S-N cars
must wait behind
the red light whilst the W-E cars are eastwards passing the crossroads.
 
Now the basic question is " {\it how should traffic engineers adjust the
value of $T_1$ in order to optimize the traffic flow through the
 intersection}?". With the assumption that the number of passengers in
each car takes
an equal average value for each direction, the optimization task is
realized by minimizing
the total waiting time of the cars per cycle of the traffic light. For
this
purpose, we introduce two quantities $N_1$ and $N_2$ which represent the
number of cars stopping (queues) at the red lights in the red phases of
the S-N
and W-E streets respectively. Clearly $N_1$ and $N_2$ are functions of
time and, in general, are divided into different lanes on the
streets. The dynamics of $N_1$ and $N_2$ are read from the following
equations in which $n$ denotes the cycle number of the traffic light:\\
\be
N_1(nT+T_1)=[N_1(nT) + (\alpha_1 - \beta_1)T_1] \theta
\ee
\be
N_2(nT+T_1)=N_2(nT) + \alpha_2T_1
\ee
\be
N_1( (n+1)T )= N_1(nT+T_1) +\alpha_1(T-T_1)
\ee
\be
N_2( (n+1)T )=[N_2(nT+T_1) + (\alpha_2  -\beta_2)(T-T_1)] \theta
\ee

The $\theta$ symbols ensure the positiveness of the quantities
in the brackets, i.e., the value of $\theta$ is one if the quantity in the
bracket is positive and zero elsewhere. This limitation is dictated to us
since, by definition, the quantities $N_1$ and $N_2$ can only take
positive values ( they denote queues' lengths). Let us investigate the
different situations in more
details. For instance in the equation $(1)$, the case $\theta =1$
which corresponds to $N_1(nT+T_1) > 0 $ 
describes the situation that after the S-N lights goes red (in
the $n$-th cycle), the whole queue of vehicles has not pass the
cross and  only a part of the queue has managed to pass 
during the green-phase. The other case, i.e., $\theta =0$ which 
corresponds to $N_1(nT+T_1)=0$ indicates that the $n$-th queue waiting in
S-N direction has completely passed the cross during the time interval $
nT \leq t \leq nT+T_1 $. The same arguments apply to eq. (4).\\  

We now define the total waiting time (TWT) of the vehicles per cycle of
the traffic light. It is the total time wasted by the vehicles during
their
stop in the red phases. The TWT is the sum of the sub-waiting
times of each direction. Denoting the TWT and the sub waiting times by
$T^w$, $T^{(w,1)}$ and $T^{(w,2)}$ respectively, the following equations
could be written for the $n$-th cycle.
\be
T_{n \rightarrow n+1}^{(w,1)}=
N_1(nT+T_1)(T-T_1) +
{1\over 2}\alpha _1 (T-T_1)^2
\ee

\be
T_{n\rightarrow n+1}^{(w,2)} = N_2(nT)T_1
+ {1\over 2} \alpha_2T_1^2
\ee

Let us consider eq.(5). The waiting time of the $n+1$-th S-N queue is
divided into two part. The first part is related to the initial length of
the queue: $N_1(nT+T_1)$. If this initial part has a non-zero length,
then
the time wasted by the initial vehicles is simply their number times the
total
period of the red phase. This leads to the first term of eq.(5). 
The second part is related to the contribution given by the new oncoming
 vehicles
arriving
at the S-N direction of the crossroads during the red period which lasts
for $T-T_1$ seconds. Since we have assumed the vehicles arrive at a
constant rate $\alpha_1$, the time wasted by the vehicles arriving in the
infinitesimal interval $[t,t+dt]$ of the red interval is simply their
number ($\alpha_1 dt$) times the remaining time to the green signal (
$T-T_1 -t$). Therefore the total contribution is given by integrating over
the red period.\\
$$ \int_0^{T-T_1}\alpha_1 dt (T-T_1-t)={1 \over 2} \alpha_1(T-T_1)^2$$  
Which is the second term on the right hand side of eq.(5). Similar
arguments lead the eq.(6).

As clearly can be seen, the analytical expression of the TWT strongly
depends on the positiveness of the queues' lengths just after the lights
go
red. In what follows, we show that different traffic status can be
identified according to the behaviour of the quantities $N_1(nT+T_1)$ and
$N_2(nT)$. Let us look at the first cycle, i.e., $n=0$. If $N_2(0)=0$ (a 
complete passing of the previous W-E queue) then it is easily seen that in
order to have a complete passing of the next W-E queue, one should have
the condition $\alpha_2 T -\beta_2 (T-T_1) \leq 0 $. In this case, one has
$N_2(T)=0$. It can be easily verified that that provided that the above
stability condition holds, we have no W-E queues after the W-E light goes
red for the general $n$-th cycle. This characterizes a {\it light traffic}
state in which the whole queue can pass the crossroads during one
green time. Similar arguments for the S-N direction shows
that provided the stability condition $\alpha_1T-\beta_1T_1 \leq 0$ holds,
we have a stable, light traffic state in the S-N direction and that
$N_1(nT+T_1)=0$ for general $n$. The case which traffic condition is light
in both
directions ( State I ) is characterized by the following
stability conditions : 

\be
\alpha_2 T -\beta_2(T-T_1) \leq 0 \; ; \; \; \; \alpha_1 T - \beta_1T_1
\leq 0
\ee

Which result in the following relations:
\be
 N_1(nT+T_1)=N_2(nT)=0 , \; \; \; n=1,2,...
\ee
In this state, The TWT is independent of the cycle number $n$.
Putting the above conditions into eqs.(5,6) and

minimizing the TWT with respect to
$T_1$ leads to the following equation:
\be
T_1= {\alpha_1 \over \alpha_1 + \alpha _2} T
\ee
Inserting the above answer in the stability
conditions (7) yields to the following constraints among the rates.\\
\be
 \beta_2 \geq \alpha_1 + \alpha_2 , \; \; \; \beta_1\geq \alpha_1 +
\alpha_2 
\ee

We now investigate a totally
different situation, i.e., a crowded crossroads in both directions. 
Let us again look at the first cycle. Supposing the first cycle
is characterized by the conditions $ N_2(0) , N_1(T_1) > 0 $. one could
easily verify that provided the following relations hold :
\be
 \alpha_1 > \beta_1 , \; \; \;  \alpha_2 T > \beta _2 (T-T_1) 
\ee
We have a stable
condition in the next cycles.
\be
N_1(nT+T_1) ,  \; \; N_2(nT) > 0 
\ee
 In sharp
contrast to the State I, in this state which is characterized by the
equation (11) and referred to as the
state II, the values of $N_1$ and $N_2$ are functions of the cycle
number. This is easily seen by the following relations:\\
\be
N_1(nT)=N_1(0) + n(\alpha_1T-\beta_1T_1)
\ee 

\be
N_2(nT)=N_2(0) + n[\alpha_2T- \beta_2(T-T_1)]
\ee

\be
N_1(nT+T_1)=N_1(0) + n(\alpha_1T-\beta_1T_1) +(\alpha_1-\beta_1)T_1
\ee

\be
N_2(nT+T_1)= N_2(0) + n[\alpha_2T-\beta_2(T-T_1)] +\alpha_2T_1
\ee

The above relations show that queues' lengths grow linearly with time and
a complete passing of a queue in one cycle is not possible. Drivers
should wait more
than one cycle in order to pass the crossroads.
It can be shown that in the large cycle-number limit, the sub-waiting
times are as follows:\\
\be
T_{n\rightarrow n+1}^{(w,1)} \sim n(T-T_1)(\alpha_1 T-\beta_1 T_1)
\ee
 and
\be
T_{n\rightarrow n+1}^{(w,2)} \sim nT_1(\alpha_2 T -\beta_2 (T-T_1) )
\ee
We now minimize the TWT with respect to $T_1$ which leads to the following
equation:
\be
T_1={ \beta_1 +\beta_2 + \alpha_1 -\alpha_2 \over 2(\beta_1 + \beta_2) } T
\ee
the consistency of this solution with stability conditions (11) yields the
following constraints:
\be
\alpha_1 >\beta_1 , \; \; \;  \alpha_1 \beta_2 +
2\alpha_2 \beta_1 +\alpha_2 \beta_2 \geq \beta_1 \beta_2 + \beta_2^2
\ee
Also positiveness of $T_1$ itself imposes the extra restriction $
\alpha_2 < \beta_1 + \beta_2 +\alpha_1 $.\\

Next we consider the situation
(state III) where in the first cycle of the traffic light, one has
$N_1(T_1)=0$ but
$N_2(0) > 0$. This corresponds to the situation in which the S-N street
has
a light traffic flow while the W-E street has a heavy one. The conditions for
a stable pattern are:
\be
\alpha_1T-\beta_1 T_1\leq 0 ,  \; \; \;   \alpha_2T-\beta_2(T-T_1) \geq 0 
\ee
The above stability conditions ensures the following relations:
\be
N_1(nT+T_1)=0, ~~~~ N_2(nT) >0
\ee
In the large cycle-number limit, minimizing of the TWT leads to the
value $T_1= { \beta_2 - \alpha_2 \over 2\beta_2 }T$. It could be easily
checked that the above solution is inconsistent with
the stability conditions, and hence, is not acceptable as an optimum
signalization of traffic light. In the region determined by the stability
conditions (21), the TWT is positive definite and therefore its minimum
coincides with the upper limit of the inequalities (21). Therefore one
finds:
\be
T_1=max( { \alpha_1 \over \beta_1} , {\beta_2 - \alpha_2 \over \beta_2} )T
\ee  

A similar argument applies to the final state 
( state IV ) which is characterized by $N_1(nT+T_1) > 0$ and
$N_2(nT)=0$ ( it is sufficient to interchange the indices one and two ).\\

%Before ending this section, we would like to mention the fact that in the
%heavy traffic condition in a road, one loses the periodic condition. For
%instance in a heavy traffic condition of the S-N street one has $
%N_1(nT+T_1) \neq N_(n+1)T + T_1)$ (queue length grows linearly with
%time). 

\section{One-Way to Two-Way Crossroads}
At this stage, we consider another 
frequent type of a crossroads. Here we let
the vehicles move in both S-N as well as N-S directions
but still the
vehicles in the W-E street are restricted to move eastwards. This situation
describes a one-way to two-way urban intersection. Consequently each
cycle
consists of three phases. In the first phase which lasts for $ 0 \leq t
\leq T_1$, the traffic light is green for the S-N cars and red for the
other two directions. During the second
phase which starts at $T_1$ and finishes at $T_2$, the traffic light is 
green for the N-S cars and red for the other two directions. In 
the final phase, which lasts for $ T_2 \leq t \leq T $, the traffic light
remains green for the W-E cars and red for the N-S as well as S-N
directions.
The entrance rates are taken to be $\alpha_1, \alpha_2 $ and $\alpha_3$
and we denote the passing rate by $\beta_1, \beta_2$ and $\beta_3$ for
each direction respectively. The starting time of the cycles is chosen
to be the moment
at which the traffic light turns green for the S-N direction. Similar
equations for the queues' lengths $N_1,N_2$ and $N_3$ could be
written down and in principle one can evaluate the TWT in terms of these
quantities. The
exact form of the TWT strongly depends on the positiveness of the
queues' lengths just after the traffic lights goes red for the respective
direction. In the case under consideration, eight different possibilities
are identified due to the traffic conditions (two possibilities for each
movement direction).
Now we write the explicit expression for the TWT during the $n\rightarrow n+1$
cycle.
\be
T_{n \rightarrow n+1}^{(w,1)} = N_1(nT+T_1)(T_2-T_1) + {1\over 2}
\alpha_1(T - T_1)^2 +N_1(nT+T_2)(T-T_2)
\ee 
\be
T_{n \rightarrow n+1}^{(w,2)} = N_2(nT)T_1 + {1\over 2} \alpha_2T_1^2 +
N_2(nT+ T_2)(T-T_2) + {1\over 2} \alpha_2 (T-T_2)^2
+ \alpha_2T_1(T-T_2)
\ee 
\be
T_{n \rightarrow n+1}^{(w,3)} = N_3(nT)T_1 + {1\over 2} \alpha_3T_2^2
+N_3(nT+T_1)(T_2-T_1)
\ee 
Let us only discuss the most probable one which
corresponds to light traffic conditions in all directions:
\be
N_1(nT+T_1) = N_2(nT+T_2) = N_3(nT) =0
\ee 
It could be
easily verified that the triple stability condition for the validity of
the above assumptions are as follows:
\be
\alpha_1 T - \beta_1 T_1 \leq 0, ~~~  \alpha_2 T -\beta_2 (T_2-T_1) \leq 0
~~~ \text{and}~~~
\alpha_3T-\beta_3(T-T_2)
\ee
Inserting eq.(27) into eqs.(24-26) and minimizing the TWT with respect to
$T_1$ and $T_2$ 
leads to the following fixation of $T_1 , T_2.$
\be
T_1 = T{ \alpha_1(\alpha_2 + \alpha_3) -\alpha_2\alpha_3 \over
 \alpha_1(\alpha_2 + \alpha_3) + \alpha_2\alpha_3 }
\ee
\be
T_2 = T{ 2\alpha_1 \alpha_2 \over
 \alpha_1(\alpha_2 + \alpha_3) + \alpha_2\alpha_3 }
\ee

One directly observes that in the symmetric case of equal arrival
rates ($\alpha_1=\alpha_2=\alpha_3$),
 $T_1$ and $T_2$ take the expected values ${T \over 3}$ and ${ 2T
\over 3}$ respectively. The other point which must be mentioned is that in this
 traffic state, $T_1$ and $T_2$ do not depend on the passing rates,
and are solely determined by the arrival rates.\\
Another extreme is the
situation when all of the three directions are carrying a heavy traffic
flow.
It could be anticipated that the stability conditions for the
positiveness
of $N_1(nT+T_1), N_2(nT+T_2)$ and $N_3(nT)$ are $\alpha_1 > \beta_1$,
$\alpha_2T_2 -\beta_2(T_2-T_1) >0$ and $\alpha_3T -\beta_3(T-T_2) >0$. In
this case, all the three sub-waiting times depend on the cycle number $n$
and it can be shown that in the large $n$ limit, the minimization of the
TWT
give rises to the following values for $T_1$ and $T_2$.
\be
T_2= {T \over 2} {2\beta_1\beta_2 + \beta_2\alpha_1 +
\beta_1\alpha_2-\alpha_3 (\beta_2 +\beta_1) + \beta_3(\beta_1+\beta_2)
\over
\beta_1\beta_2 + \beta_2\beta_3 + \beta_3\beta_1 }
\ee 
\be
T_1= {T \over 2} {\beta_1\beta_2 + \beta_3 \beta_1 + \beta_2\alpha_1 +
\beta_3\alpha_1- \alpha_2 \beta_3 - \beta_2\alpha_3 \over
\beta_1\beta_2 + \beta_2\beta_3 + \beta_3\beta_1 }
\ee 
Here in contrast to (29) and (30) , the passing rates
appear in the expressions of $T_1$ and $T_2$ and it can be seen that in
the fully symmetric condition, one again obtains that $T_1$ and $T_2$ are
one-third and two-thirds of $T$ respectively.\\

\section{ Empirical Data }

For comparison of our model to the empirical data, a time-series analysis
on two of Tehran intersections were carried out. The central part of
Tehran is under control of the SCATS ({\it Sydney Co-ordinated Adaptive
Traffic System}) \cite{scats1, scats2} that is an intelligent
traffic controller. The strategy followed by most of urban traffic
control systems is based on establishing green-waves along the major
streets of cities. One popular strategy consists of dividing the city
intersections into different sets. Each set has a leading mother
crossroads (the prime crossroads in the set) and a lot of offspring
crossroads.
Each set is linked to the neibouring ones. The signalization of the
 crossroads network is determined by the implementation of green wave
between adjacent sets. The allocated green times at a crossroads is
proportional to the number of vehicles (traffic volume) approaching to the
crossroads. Since in general, there is a natural fluctuation in the
traffic volume, the amount of green times and hence the complete cycle
time of traffic lights are variables in an intelligent control
method.\\

We considered two different intersections. The first one which
is located in Tehran downtown connects Valiasr St. to Takhtejamshid St.
Both of these  are one-way and major streets. The other crossroads
connects Abbasabad St.(one-way) to Mahnaz St.(one-way) . In contrast to
the previous case, here the first street is a major while the second
street is a minor one. The data set is provided by magnetic counting loops
which
are installed just before the pedestrian-lines of each crossroads. The
data
were collected on second of July 2000. 
The data consist of the sub-phase green times and the numbers of vehicles
passed during the green times for each cycle of the traffic lights
(traffic volumes). In
our major-to-minor crossroads and for each cycle of the traffic light, we
evaluated the ratio of the major street green-time to the total cycle
time. We call this quantity the {\it time-cycle-ratio }.  
Similarly can can consider the number of passed vehicles, i.e., traffic
volume in a cycle and
introduce the {\it volume-cycle-ratio } which for each cycle is
obtained by dividing the number of passed vehicles during a sub-phase
green time to the whole number of passed vehicles during a complete cycle.
Figures 1-3 belong to the major-to-minor crossroads. As seen from the
graphs, the time-cycle-ratio allocated to
the major street strongly fluctuates due to demand fluctuations received
by the crossroads. In order to have a rough estimation of our model
parameters, we considered a two-hour time interval between
12:30-14:30 during which we have the least fluctuations in the traffic
volume. It is empirically observed that in this two-hour period, the
traffic state is light and the queues are cleared
during one cycle. Therefore,
the optimal green times should be evaluated from eq.(9). According to this
equation, we only need to know the
the ratio of ${ \alpha_1 \over \alpha_2}$.

We should mention that
due to the position of the magnetic counting loops, they are unable to
measure the upstream fluxes which are directly related to the parameters
$\alpha_1$ and $\alpha_2$. For a better estimation of the in-flow
parameters, one should install another set of magnetic loops a few meters
upwards the pedestrian-lines.
We approximated the ratio of ${ \alpha_1 \over \alpha_2}$ by
the ratio of the traffic volume which has passed through
the major street to the traffic volume of the minor one during the
two-hour period. The two-hour traffic volumes are simply obtained by
adding the cycle volumes together. 
This yields the value
${\alpha_1 \over \alpha_2 }= 0.35 $. Putting this value into
equation (9) yields ${T_1 \over T}=0.74$. On the other hand, the averaged
value of the empirical time-cycle-ratio of the major street over the
two-hour period leads to
the result ${T_1 \over T }=0.64$ which differ by ten percent from the
value predicted by the theory. Figures 4-6 belong to our
major-to-major crossroads. Here we focused on the interval 13:30 - 15:30.
Analogous to the major-to-minor, the traffic state is light.
The empirical averaged value of
${T_1 \over T}$ is 0.52 while the same amount evaluated from eq. (9)
is 0.56 ($T_1$ refers to the green time of Takht. Street). Here we
observe that the difference between the fixed time method
(theory) and real-time method (intelligent control) is less than one
obtained in the major-to-minor crossroads. As depicted from the diagrams,
in the major-minor crossroads, we observe more fluctuations in the
time-cycle-ratio in comparison with the major-major crossroads. These
fluctuations are enhanced in the volume-cycle-ratio. The
least fluctuation belongs to the time-cycle-ratio of the major-major
crossroads. Comparing the averaged values of these time-cycle-ratio
values over certain intervals leads to a better understanding of the
traffic state.

\section{Summary and Conclusion}
In conclusion, we have developed a
prescription for the traffic-light
programming at a single urban crossroads. The method is based on minimizing the
total waiting-time of cars stopping in the red phases of the traffic
light. In our model the total period of the cycle is assumed to be a fixed
value. We have also assumed that vehicles arrive at the crossroads with
constant
time rates. This is equivalent to a constant time-headway between cars. In
reality we have a fluctuating time-headway due to the natural
fluctuation in the traffic volume. As a first stage of an analytical
treatment, we have taken the arrival rates to be constant in time.
However, for a more realistic description, one should remove this
restriction and assume that the time headway 
satisfies a random distribution function. Work along this assumption is
in progress. The other point concerns the passing-rates.
One should note that throughout the paper, the passing-rate of cars
from the crossroads are taken to be constants. This is valid only if the
green-phase time is not so long such that the time-headways between the
cars exceed certain values. 
The value of $T$ should be so tuned that during the green phases,
time-headway is less than a certain value.
In fact, in the model, the values of the passing rate refer to the maximum
capacity of cross ( maximum number of cars passing the cross in the unit
of time ), which is plausible if the $T$ is appropriately adjusted with 
the congestion of the crossroads. The empirical value of
the passing rates are determined by the crossroads
characteristics such as road conditions, number of lanes, speed limits
etc. Our model is more appropriate for rush ours during which the
time-headways are minimum and the crossroads are operating with their
maximum capacities. In the following table, we have summarized our
theoretical optimized green times including stability conditions in four
different traffic states of the one-way to one-way crossroads.\\
 \begin{center}
{\large Table}
\end{center}
    
\begin{tabular}{cccc}

 Traffic state  & Optimized green time & Stability conditions &  \\
 \hline
  &  &  &  \\
State I (S-N light, W-E light)  &
$T_1={\alpha_1 \over \alpha_1 + \alpha_2 }T $ &
 $ \beta_1 , \beta_2 \geq \alpha_1 + \alpha_2,~~   $ &\\ 
 \hline   
 &  &  & \\ 
State II (S-N heavy, W-E heavy)  &
  $T_1={\beta_1 + \beta_2+ \alpha_1 -\alpha_2 \over 2(\beta_1 + \beta_2)
}T$ &
 $ \alpha_1 \geq \beta_1, ~~ \alpha_1 \beta_2 + 2\alpha_2 \beta_1
+ \alpha_2 \beta_2 \geq \beta_1 \beta_2 + \beta_2^2 $ &
          \\ 
 \hline   
 &  &  &  \\
State III (S-N light, W-E heavy)  &
  $T_1=max ({\alpha_1 \over \beta_1 }, {\beta_2-\alpha_2 \over \beta_2})T$
&
 $ \alpha_1 \leq \beta_1,~~ \beta_2 \geq \alpha_2 $ &\\ 
 \hline
 &  &  &  \\
State IV (S-N heavy, W-E light)  &
  $T_1=max ({\alpha_2 \over \beta_2 }, {\beta_1-\alpha_1 \over \beta_1})T$
&
 $ \alpha_2 \leq \beta_2,~~ \beta_1 \geq \alpha_1 $ &\\
 \hline                    

 \end{tabular}

Optimizing the traffic at each crossroads is the
stating point of the more comprehensive problem of city network.
Nevertheless, our model best suits those marginal intersections 
of cities where the effect of the other crossroads is suppressed.\\

\section{ Acknowledgments}
 M.E.F is grateful to {\it Tehran Traffic
Control
Company } and in particular to F. Zolfaghari for the data support. We
highly appreciate V. Karimipour for
fruitful comments and critically reading the manuscript and would like to
express our gratitude to
R. Sorfleet, R. Gerami, M. Arab salmani, M. Salmani and M. Khoshechin for
useful
helps. 
\bibliographystyle{unsrt}

\begin{thebibliography}{99}

%\bibitem{tgf95} D.E.\ Wolf, M.\ Schreckenberg and A.\ Bachem (eds) {\em
%Traffic and granular flow} (World Scientific, Singapore, 1996).


\bibitem{css99}
Chowdhury, D., Santen, L., and Schadschneider, A.,
{\em Physics Reports}, {\bf 329}, 199 (2000).


\bibitem{tgf97} D.E\ Wolf and M.\ Schreckenberg (eds.) {\em Traffic and
granular flow} (Springer, Singapore, 1998).



\bibitem{tgf99} \ H.J.\ Herrmann, D.\ Helbing, M.\
Schreckenberg and D.E.\ Wolf (eds.) {\em Traffic and Granular flow}
 (Springer, Berlin, 2000).
    



\bibitem{helbbook} D.\ Helbing, {\em Vehrkersdynamik: Neue Physikalische
Modellierungskonzepte}, Springer, Berlin,1997.


\bibitem{zia} B. Schmittmann and R.K.P. Zia in : {\em Phase Transition and
Critical Phenomena }, edited by C. Domb and J.L. Lebowitz, vol. 17
(Academic Press, New York, 1995). 



\bibitem{schutzbook} G.M. Sch\"utz, in {\em Phase Transition and Critical
Phenomena }, edited by C. Domb and J. Lebowitz (Academic Press, London, in
press).



\bibitem{privbook} {\em Non-Equilibrium Statistical Mechanics in One
Dimension}, edited by V. Privman (Cambridge University Press, Cambridge,
England, 1997).




\bibitem{book}
 {\em Traffic Flow Fundamentals}, Prentice Hall (1990) by A.D.\
May. {\em Transportation and Traffic Theory}, Elsevier (1993) by C.F
\ Daganzo. 



\bibitem{bell} M.G. Bell, {\it Future Directions in Traffic Signal
Control}, {\em Transport Research: A}, {\bf 26}, no 4, 303 (1992).




\bibitem{robertson} D.I. Robertson and R.D. Bretherton, {\it Optimizing
networks of traffic signals in real-time: the SCOOT method }, {\em IEEE
Transportation on Vehicular Technology}, {\bf 40}, 11 (1991).




\bibitem{bml} O.\ Biham, A.\ Middleton and D.\ Levine, Phys.Rev. {\bf A
}{\bf 46}, R6124 (1992).





\bibitem{nagatani1} T. Nagatani, {\em J. Phys. Soc. Japan} {\bf 62}, 2656
(1993).



\bibitem{chung} K.H. Chung, P.M. Hui and G.Q. Gu, {\em Phys. Rev. E} {\bf
51}, 772 (1995).


\bibitem{cuesta} J.A. Cuesta, F.C. Martines, J.M. Molera and A. Sanchez,
{\em Phys. Rev. E} {\bf 48}, R4175 (1993).



\bibitem{nagatani2} T. Nagatani, {\em J. Phys. Soc. Japan} {\bf 63}, 1228
(1994). 



\bibitem{torok} J. T\"or\"ok and J. Kertesz, {\em Physica A} {\bf 231},
515 (1996).



\bibitem{horiguchi} T. Horiguchi and T. Sakakibara, {\em Physica A} {\bf
252}, 388 (1998); {\em Interdisc. Inf. Sci.} {\bf 4}, 39 (1998). 



\bibitem{freund} J. Freund and T. P\"oschel, {\em Physica A} {\bf 219}, 95
(1995).



\bibitem{chopard} B. Chopard, P.O Luthi and P.A. Queloz, {\em J. Phys. A}
{\bf 29}, 2325 (1996).


\bibitem{schad-deb} D.\ Chowdhury and A.\ Schadschneider, {\em Phys. Rev. 
E} {\bf 59 },
R 1311 (1999).



\bibitem{NS} K.Nagel, M.Schreckenberg, {\em J.Phys.I France} {\bf 2}, 2221
(1992).


\bibitem{ito} M. Schreckenberg, A. Schadschneider, K. Nagel and N. Ito,
{\em Phys. Rev. E} {\bf 51}, 2939 (1995).



\bibitem{brockfeld} Elmar Brockfeld, {\em Simulation von Stadtverkehr
mittels Zellularautomaten} (Diplomarbeit), University of Osnabr\"uck,
2000.



\bibitem{scats1} P. Lowrie, {\it SCATS: A Traffic Responsive Method for
Controlling Urban Traffic}, tech. rep., {\em Road and Traffic Authority},
NSW, Australia.


\bibitem{scats2} A.G. Sims, "SCATS: the Sydney co-ordinated adaptive
system", in the proceeding of the {\it Engineering Foundation Conference
on Research Priorities in Computer Control of Urban Traffic Systems}, 12,
1979.
 

\end{thebibliography}

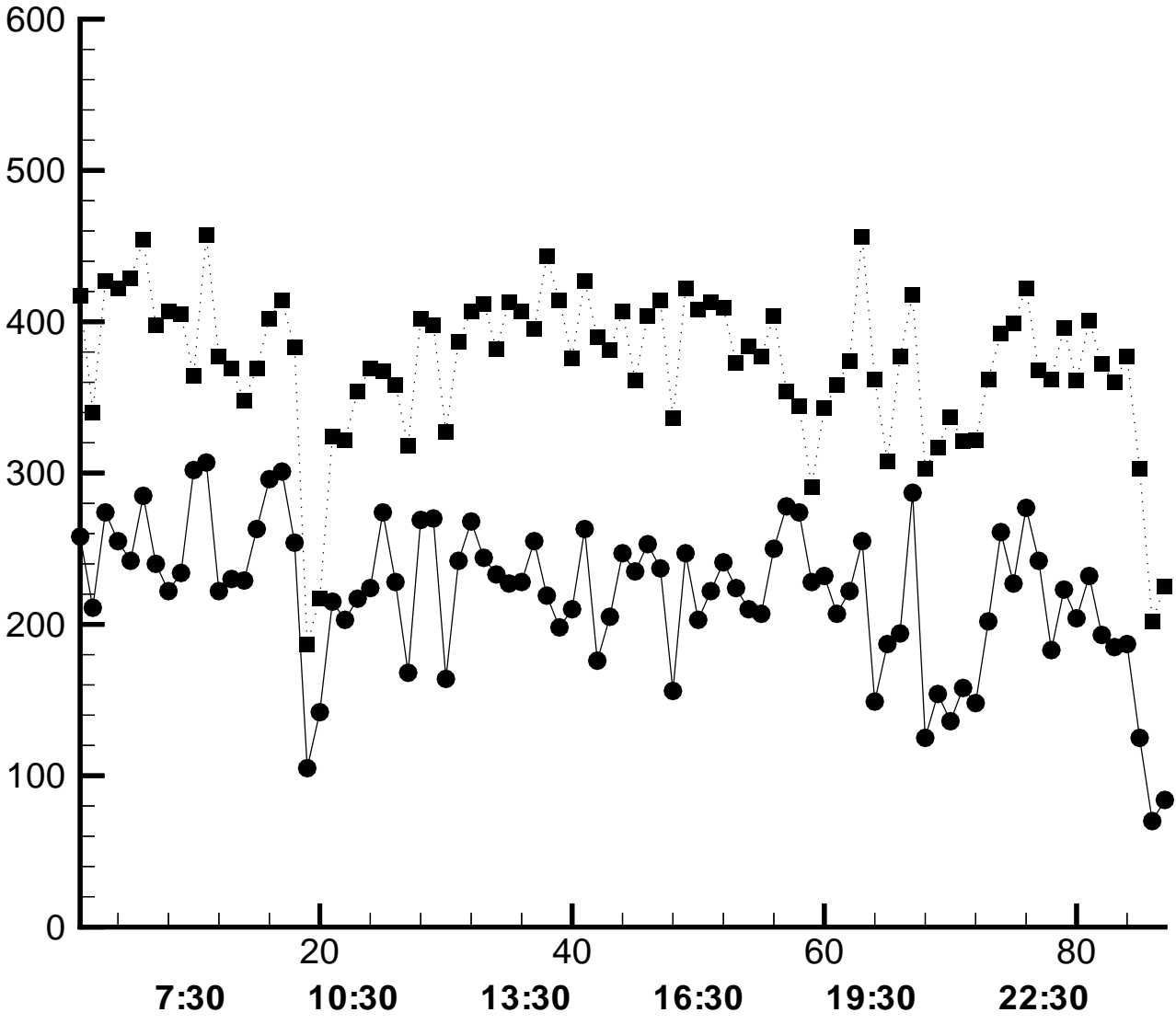
\end{document}